\documentclass[12pt]{emulateapj}
\usepackage{graphicx}
\usepackage{amsmath}
\usepackage{amssymb}
\usepackage{eucal}

\newcommand{\swift}{\textit{Swift}}
\newcommand{\fermi}{\textit{Fermi}}

\newcommand{\cm}[1]{~cm$^{#1}$}

\newcommand{\cts}{~cts\,s$^{-1}$}
\newcommand{\e}[1]{10$^{#1}$}
\newcommand{\ee}[1]{$\times$10$^{#1}$}
\newcommand{\ergs}{~ergs\,cm$^{-2}$\,s$^{-1}$}
\newcommand{\lergs}{~ergs\,s$^{-1}$}

\newcommand{\msun}{M$_{\odot}$}
\newcommand{\nh}{N$_{\rm H}$}

\shorttitle{Broadband~study of GRB~091127}
\shortauthors{E. Troja et al.}

\begin{document}
\title{Broadband~study of GRB~091127: a sub-energetic burst at higher redshift?}

\author{E.~Troja\altaffilmark{1,2}, 
T.~Sakamoto\altaffilmark{2}, 
C.~Guidorzi\altaffilmark{3,4}, 
J.~P.~Norris\altaffilmark{5}, 
A.~Panaitescu\altaffilmark{6},
S.~Kobayashi\altaffilmark{4}, 
N.~Omodei\altaffilmark{7},
J.~C. Brown\altaffilmark{2},
D.~N. Burrows\altaffilmark{8},
P.~A. Evans\altaffilmark{9},
N.~Gehrels\altaffilmark{2},
F.~E.~Marshall\altaffilmark{2}, 
N.~Mawson\altaffilmark{4},
A.~Melandri\altaffilmark{10,4}
C.~G.~Mundell\altaffilmark{4},
S.~R.~Oates\altaffilmark{11},  
V.~Pal'shin\altaffilmark{12}, 
R.~D.~Preece\altaffilmark{13}, 
J.~L.~Racusin\altaffilmark{2}, 
I.~A.~Steele\altaffilmark{4},
N.~R.~Tanvir\altaffilmark{9},
V.~Vasileiou\altaffilmark{14},
C.~Wilson-Hodge\altaffilmark{15},
K.~Yamaoka\altaffilmark{16}}

\altaffiltext{1}{NASA Postdoctoral Program Fellow}		
\altaffiltext{2}{NASA, Goddard Space Flight Center, Greenbelt, MD 20771, USA}
\altaffiltext{3}{Physics Department, University of Ferrara, via Saragat 1, I-44122, Ferrara, Italy}
\altaffiltext{4}{Astrophysics Research Institute, Liverpool John Moores University, Twelve Quays House, Egerton Wharf, CH41 1LD, Birkenhead, UK}
\altaffiltext{5}{Physics Department, Boise State University, 1910 University Drive, Boise, ID 83725, USA}
\altaffiltext{6}{Space Science and Applications, MS D466, Los Alamos National Laboratory, Los Alamos, NM 87545, USA}
\altaffiltext{7}{W. W. Hansen Experimental Physics Laboratory, Kavli Institute for Particle Astrophysics and Cosmology, Department of Physics and SLAC National Accelerator Laboratory, Stanford University, Stanford, CA 94305, USA}
\altaffiltext{8}{Department of Astronomy and Astrophysics, Pennsylvania State University, 525 Davey Lab, University Park, PA 16802, USA}
\altaffiltext{9}{X-ray and Observational Astronomy Group, Department of Physics and Astronomy, University of Leicester, LE1 7RH, UK}
\altaffiltext{10}{INAF-OAB, via Bianchi 46, I-23807 Merate (LC), Italy}
\altaffiltext{11}{Mullard Space Science Laboratory, University College London, Holmbury St. Mary, Dorking Surrey, RH5 6NT, UK}
\altaffiltext{12}{Ioffe Physico-Technical Institute, Laboratory for Experimental Astrophysics, 26 Polytekhnicheskaya, St Petersburg 194021, Russian Federation}
\altaffiltext{13}{University of Alabama in Huntsville, NSSTC, 320 Sparkman Drive, Huntsville, AL 35805, USA}
\altaffiltext{14}{Laboratoire Univers et Particules de Montpellier, Universit\'e Montpellier 2, and
CNRS/IN2P3, Montpellier, France}
\altaffiltext{15}{Institute of Astro and Particle Physics, University Innsbruck, Technikerstrasse 25, 6176 Innsbruck, Austria}
\altaffiltext{16}{Department of Physics and Mathematics, Aoyama Gakuin University, 5-10-1 Fuchinobe, Chuo-ku, Sagamihara, Kanagawa 252-5258}
\begin{abstract}

GRB~091127 is a bright gamma-ray burst (GRB) detected by \swift~at a redshift $z$=0.49 
and associated with SN~2009nz. 
We present the broadband analysis of the GRB prompt and afterglow emission
and study its high-energy properties in the context of the GRB/SN association.
While the high luminosity of the prompt emission and standard afterglow behavior 
are typical of cosmological long GRBs, its low energy release 
($E_{\gamma}$$<$3$\times$10$^{49}$~erg), soft spectrum and unusual spectral lag
connect this GRB to the class of sub-energetic bursts.
We discuss the suppression of high-energy emission in this burst, 
and investigate whether this behavior could be connected with 
the sub-energetic nature of the explosion.
\end{abstract}


\section{Introduction}\label{sec:intro}

It is well established that (most) long duration GRBs are linked to the gravitational collapse of massive stars \citep{wb06}. 
Such a connection is supported by several lines of evidence \citep[][and references therein]{hbloom11}.  
In a few remarkable cases the spectroscopic identification of a broad line Type Ic SN,
co-spatial and coeval with the GRB, provided a direct proof of the physical association between the two 
phenomena.

With the exception of GRB~030329, whose properties are roughly similar to typical long GRBs \citep{berger03}, 
GRBs with spectroscopically confirmed SNe show a peculiar behavior, 
both in their prompt and afterglow emission phases \citep{kaneko07,starling10}.
These bursts are characterized by a relatively softer spectrum ($E_{\rm pk}\lesssim$120\,keV), and a
lower energy output (E$_{\gamma,\rm iso}$$\sim$\e{48}-\e{50}\,erg) 
than standard GRBs.
They do not strictly follow the lag-luminosity relation \citep{norris02}, whereas
they generally agree with the Amati relation \citep{amati07},
but with GRB~980425 being a notable outlier.
Sub-energetic nearby bursts tend to show a faint afterglow emission, both in X-rays and 
in the optical band.  
Late time radio monitoring of their afterglows showed evidence of a quasi-spherical and 
only mildly relativistic ($\Gamma$\,$\approx$\,2) outflow \citep{sode06},
very different from the highly relativistic and collimated jets observed in long GRBs \citep{bloom03,molinari07,cenko10}.
For these reasons it has been speculated
that sub-energetic events belong to an intrinsically distinct population of bursts 
which dominate 
the local ($z\lesssim 0.5$) rate of observed events \citep{liang07,chapman07}. 

Whereas the case for spectroscopically confirmed SNe remains confined to nearby GRBs, 
at higher redshifts ($0.3<z<1$) the emergence of the associated SN is pinpointed
by a late-time optical rebrightening or ``bump" in the 
afterglow  light curves \citep{bloom99,zeh05,tanvir09}. 
Though alternative explanations for such a feature are plausible \citep{esin00,waxman00},
a spectroscopic analysis of some of these SN bumps
supports their similarity with bright Type Ic SNe \citep[e.g.][]{dellavalle06,sparre11}.
This is the case of GRB~091127, detected by the \swift~satellite \citep{gehrels04} 
at a redshift of $z=$0.49, and associated with SN2009nz.
\citet{cobb10} identified in the GRB afterglow a late-time optical rebrightening, 
peaking at a magnitude of $I$\,=\,22.3$\pm$0.2\,mag at $\sim$22~d after the burst, and attributed it to the SN light.
The photometric properties of SN2009nz resemble SN1998bw \citep{galama98}, 
though displaying a faster temporal evolution and a slightly dimmer peak magnitude. 
More recently, the spectroscopic analysis presented by \citet{berger11}
uncovered the typical undulations of broad line Type~Ic SNe associated with nearby GRBs, thus confirming the SN origin of the photometric bump.
\citet{berger11} concluded that the explosion properties of SN2009nz
(E$_K$$\approx$2\ee{51}\,erg, 
M$_{ej}$$\sim$1.4\,\msun, and M$_{\rm Ni}$$\approx$0.35\,\msun)
are remarkably similar to SN2006aj \citep{pian06}, associated with GRB~060218. 
GRB~091127 therefore represents one of the best cases linking 
long GRBs and SNe at redshifts $z$$>$0.3.

While previous works mainly focused on the properties of SN2009nz 
and its environment \citep{cobb10,vergani11}, 
in this paper we present a broadband analysis of the GRB prompt and afterglow
emission and study the high-energy properties of the explosion in the 
context of GRB/SN associations.
Being a bright and relatively nearby burst, GRB~091127 has a rich multi-wavelength coverage up to very late times, 
which allows us to study in detail its spectral and temporal evolution  \citep[see also][]{filgas11} and compare it to other well-known cases of GRBs/SNe.

The paper is organized as follows:
our observations are detailed in $\S$~\ref{sec:obs}.
In $\S$~\ref{sec:data} we present a multi-wavelength timing and spectral analysis
of both the prompt and the afterglow emission; 
our results are presented in \S~\ref{sec:res}
and discussed in $\S$~\ref{sec:discuss}.
Finally, in $\S$~\ref{sec:end} we summarize our findings and conclusions.
Throughout the paper, times are given relative
to the \swift~trigger time T$_0$, t=T-T$_0$,
and the convention $f_{\nu,t} \propto \nu^{-\beta}t^{-\alpha}$
has been followed, where the energy index $\beta$ is related to the
photon index $\Gamma$\,=\,$\beta+1$.
The phenomenology of the burst is presented in the observer's time
frame. Unless otherwise stated, 
all the quoted errors are given at 90\% confidence level for one interesting
parameter \citep{lampton76}.\\

\section{Observations and Data Reduction}\label{sec:obs}

GRB~091127 triggered the \swift~Burst Alert Telescope (BAT; \citealt{bat05}) at 23:25:45 UT on 2009 November 27
\citep{swift}.
It was also observed by {\it Konus-Wind}, 
{\it Suzaku} Wide-band All-sky Monitor (WAM), 
and the \fermi~Gamma-Ray Burst Monitor (GBM). 
The burst was within the field of view of the \fermi~Large Area Telescope 
(LAT; \citealt{atwood09}), at an angle of 25$^{\circ}$ from the boresight.

The 2-m Liverpool Telescope (LT) responded robotically to the {\it Swift} alert 
and began observing at 23:28:06~UT, 141 s after the BAT trigger.
The detection mode of the automatic LT GRB pipeline \citep{guidorzi06}
identified a bright optical afterglow \mbox{($r^{\prime}=15.4$~mag)}
at $\alpha=02^{\rm h} 26^{\rm m} 19\fs89$, $\delta=-18^\circ 57\arcmin 08\farcs6$ (J2000) (uncertainty of $0\farcs5$; \citealt{gcnliverpool}).
Observations were obtained with $r^{\prime}i^{\prime}z^{\prime}$ filters until $2.3$~hours post burst.
The afterglow was monitored with both the Faulkes Telescope South (FTS) and LT up to 6 days post-burst
within the $BVRr^{\prime}i^{\prime}$ filters.
Magnitudes of field stars in $BVR$ were calibrated using Landolt standard
stars \citep{landolt92} obtained during following photometric nights.
SDSS $r^{\prime}i^{\prime}z^{\prime}$ magnitudes of the same field stars were obtained using
the transformations by \citet{jordi06}.
Early time observations were also obtained using SkycamZ, 
mounted on the LT tube.
Observations are filter-less (white light) to maximize
the throughput of the optics.  The data were dark and bias subtracted in
the usual fashion and flat fielded using a stack of twilight exposures.
Standard aperture photometry was carried out using two local
reference stars, and calibrated by comparison with R band frames 
of the same field.

Due to an Earth limb constraint, \swift~did not immediately slew to the burst
location and follow-up observations with its two narrow field instruments, 
 the X-Ray Telescope \citep[XRT;][]{xrt05} and the Ultra-Violet Optical Telescope \citep[UVOT;][]{uvot05}, began 53 min after the trigger. 
%
As the X-ray afterglow was still bright ($\sim$10\cts), XRT started collecting data in  Windowed Timing (WT) mode, and automatically switched to Photon Counting (PC) mode when the source decreased to $\lesssim$2\,\cts.
Follow-up observations monitored the X-ray afterglow for 36~d for a total net exposure of 760 s in WT mode and 470 ks in PC mode.
The optical afterglow was detected by UVOT in the White, $v$, $u$, $uvw1$, 
and $uvm2$ filters at a position consistent with the LT localization.
The detection in the UV filters is consistent with the low redshift
$z$=0.49 of this burst.
\swift/XRT and UVOT data were reduced using the HEASOFT\footnote{http://heasarc.gsfc.nasa.gov/docs/software/lheasoft/} 
(v6.11) and \swift~software (v3.8) tools and latest calibration products. 
We refer the reader to \citet{evans07} for further details on the XRT data
reduction and analysis. 
The UVOT photometry was done following the methods
described in \citet{breeveld10} with adjustments to 
compensate for the contamination of a nearby star.

In order to monitor the late time X-ray afterglow, 
two Target of Opportunity observations were performed
by the {\it Chandra X-Ray Observatory} 
at t=98~d for a total exposure of 38~ks and t=188~d
for a total exposure of 80~ks. 
{\it Chandra} data were reduced 
using version 4.2 of the CIAO software. 
Source events were extracted from a 2 pixel radius
region around the GRB position, while the background
was estimated from a source-free area using a 20 pixel radius region.
\\

\section{Data analysis}\label{sec:data}

 \subsection{Gamma-ray data}\label{sec:gamma}
\subsubsection{Temporal analysis}\label{sec:time}

\begin{figure}[!t]
\centering
\vspace{0.2cm}
\includegraphics[scale=0.63]{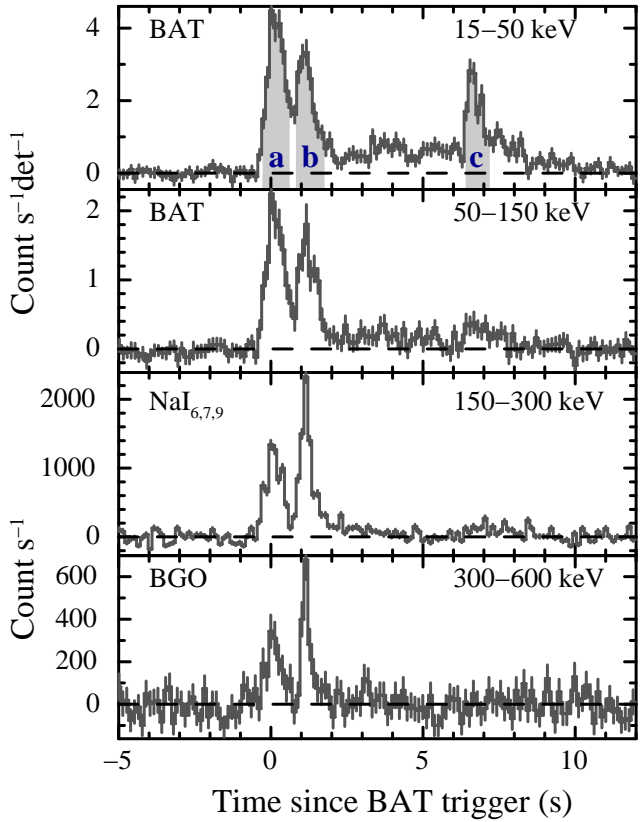}
\caption{\swift/BAT (top two panels) and \fermi/GBM (bottom
two panels) background-subtracted light curves of GRB~091127 with a 128 ms binning. 
The grey areas labelled as $a$, $b$ and $c$ in the top panel 
show the three intervals selected for the time-resolved spectral analysis. 
Error bars are 1 $\sigma$.}
\vspace{0.2cm}
\label{fig:batlc}
\end{figure}

Figure~\ref{fig:batlc} presents the prompt emission light curves
with a 128 ms time resolution and in four different energy bands.
The burst duration, defined as the interval containing  
90\% of the total observed fluence, is $T_{90}$ (15-350~keV)=7.1$\pm$0.2~s.
The burst temporal profile is characterized by two main peaks, 
at t$\sim$0~s and t$\sim$1.1~s, respectively. They 
are clearly detected up to $\sim$600 keV and display a 
soft-to-hard spectral evolution. 
A period of faint, spectrally soft emission lasting $\sim$8\,s, follows.
On top of it a third peak at ~t$\sim$7~s is visible at 
energies below 50 keV.  


Spectral lags were calculated by
cross-correlating the light curves in the standard 
BAT channels: 1 (15-25~keV), 2 (25-50~keV), 
3 (50-100~keV), 4 (100-350~keV).
In order to increase the signal-to-noise in the higher energy channels, 
the analysis was performed on non mask-weighted lightcurves, each with 
a 8~ms time resolution.
We derived $\tau_{31}=$2.2$^{+22.8}_{-11.3}$\,ms and $\tau_{42}=$$-$9.2$^{+8.2}_{-6.5}$\,ms, where the quoted uncertainties (at a 1\,$\sigma$ confidence level) were evaluated by simulations. 
Lag analysis reveals a significant difference between the two main $\gamma$-ray peaks ($a$ and $b$ in Fig.~1). The former shows positive lags, $\tau_{31}=$36$^{+24}_{-16}$\,ms and $\tau_{42}=$16$^{+13}_{-13}$\,ms, while
the latter has negligible or negative lags, $\tau_{31}=$$-$2$^{+12}_{-12}$\,ms and $\tau_{42}=$$-$14$^{+7}_{-3}$\,ms.

\subsubsection{Search for high-energy $\gamma$-ray emission}\label{sec:lat}

The {\it Fermi}/LAT data were searched for emission during the prompt $\gamma$-ray phase and over longer timescales (up to 10 ks).
The searches were performed by means of an unbinned likelihood analysis \citep{latref}.
We used the Pass7V6 Transient class events with a reconstructed energy above 100 MeV.
We selected events within 12 degrees around the best burst position (see~\S\ref{sec:obs}), 
and applied a cut on zenith angle at 105$^{\circ}$ in order to limit the contamination from the bright Earth's limb.
For the Transient data class the dominant background component is the isotropic background due to residual charged particles misclassified 
as $\gamma$-rays. 
We modeled it by using the tool developed by the LAT collaboration that can predict the hadronic cosmic ray and  \mbox{$\gamma$-ray} components of the background with an accuracy of $\approx$10-15\%
\citep{latref}.  
We also added the template {\tt gal\_2yearp7v6\_v0.fits}\footnote{ Available at the Fermi Science Support Center web site  http://fermi.gsfc.nasa.gov/ssc/data/access/lat/BackgroundModels.html}
 describing the Galactic diffuse emission due to the interaction of cosmic rays with the gas and the interstellar radiation field.

No significant excess above background was found.
Following the procedure described in \citet{latref} and by  
fixing the photon index to 2.25,
we derived a 95\% upper limit of 2.8\ee{-8}\,erg\,cm$^{-2}$\,s$^{-1}$ in the 100~MeV-1~GeV energy range and of 1.6\ee{-8}\,erg\,cm$^{-2}$\,s$^{-1}$ in \mbox{the 1~GeV-10~GeV} energy range during the prompt emission  interval \mbox{(-0.3\,s$<$t$<$8.2\,s). }

\subsubsection{Spectral analysis}\label{sec:spectra}
We performed a time-averaged 
and a time-resolved spectral analysis, 
selecting the time intervals in correspondence of the
three main pulses as shown in Figure~\ref{fig:batlc} (top panel).
The spectral fits were performed in the 15-150 keV
energy band for BAT, 20~keV-10~MeV for {\it Konus-Wind}, and 120~keV-3~MeV for {\it Suzaku}-WAM.
Following Sakamoto et al. (2010) we added a 5\% systematic error
in the WAM spectra below 400 keV.
The intercalibration between BAT, {\it Konus-Wind} and {\it Suzaku}-WAM was
extensively studied by Sakamoto et al. (2010), showing
an overall agreement in the effective area correction 
($<$20\%) between the three instruments.
GBM data were fit in the 8-860 keV band for the NaI detectors, 
and in the 200~keV-40~MeV for the BGO detector. 
Given the brightness of this burst
we added a 5\% systematic error to the GBM data, needed
to improve the fit acceptance of the time-averaged analysis.
A cross-calibration study has not been performed with the \fermi~data yet.
Previous works \citep[e.g.][]{page09} report a typical effective area correction factor of $\sim$1.23 compared to a value of unity for BAT, 
and in our analysis we found consistent values.

The best fit spectral parameters were estimated 
using the maximum likelihood method and, when necessary, 
by applying different statistics to the data. 
BAT mask-weighted spectra have Gaussian distributed uncertainties, 
and they require the $\chi^2$ statistics to be applied. 
LAT spectra are instead characterized by low counts, and they can only be modeled
using the Poisson distribution. In order to properly account for the Poissonian
nature of the source counts and for the Gaussian uncertainties associated to 
the LAT background model \citep{latref}, 
we used the profile likelihood statistic as implemented in the
option PGSTAT of XSPEC \citep{arnaud11}. 
Table~\ref{tabpha1} reports the spectral fit results for the
time-averaged analysis.
Different spectral models, usually adopted
to describe the GRB prompt emission spectrum, were fit to the data: 
a power-law (PL), a power-law with a high-energy cut-off (CPL; $F(E) \propto E^{\alpha} e^{-E/E_{\rm cut}}$), 
a Band model \citep{band93}, and a Band model with a 
high energy cut-off (Band+Cut). 
We also included the log-parabolic function (LOGP; $F(E) \propto E^{\alpha+\beta {\rm log }E}$)
suggested by \citet{massaro10}.
The last column of Table~~\ref{tabpha1} reports the fit statistics (STAT) 
and degrees of freedom (d.o.f.) for each model. 
In general STAT=$\chi^2$, when LAT data 
were included in the fit STAT=$\chi^2$+PGSTAT.

\begin{table*}[!thb] 
\centering
\caption{Spectral Fit Results of the time-averaged analysis}
\begin{tabular}{ccccccc}
\hline
\ \ \ \ \ \ Detector \ \ \ \ \ & 
\ \ \ \ \ \ Model \ \ \ \ \ \ &
\ \ \ \ \ \ \ \ \ $-$$\alpha$\ \ \ \ \ \ \ \ &
 \ \ \ \ \ \ \ \ \ $-$$\beta$\ \ \ \ \ \ \ \ & 
 \ \ \ \ \ \ E$_{\rm pk}$ (keV)  \ \ \ \ \ \ & 
 \ \ \ \ \ \ \ E$_{\rm cut}$ (keV) \ \ \ \ \ & 
 \ \ \ \ STAT/d.o.f\ \ \ \ \\
\hline
BAT & PL & 2.17$\pm$0.07 & -- & -- & -- & 52/57 (0.91) \\
KW & PL & 2.19$\pm$0.04 &  -- & -- & -- & 74/59 (1.26) \\
KW & CPL & 2.00$\pm$0.10 &  -- & -- & 510$^{+440}_{-180}$ & 57/58 (0.98) \\
WAM & PL & 2.35$^{+0.17}_{-0.19}$ &  -- & -- & -- & 23/25 (0.90) \\
WAM & CPL & 1.9$\pm$0.5 &  -- & -- & $>$400 &20/24 (0.85) \\
GBM & LOGP & 0.73$\pm$0.14 & 0.37$\pm$0.04 & -- & -- & 465/396 (1.18) \\
GBM & Band & 1.20$\pm$0.16 & 2.23$\pm$0.04 & 39$\pm$5 & -- &  457/395 (1.16) \\
GBM & Band+Cut & 0.3$^{+0.9}_{-1.3}$ & 1.94$\pm$0.08 & 25$^{+16}_{-5}$ & 500$^{+300}_{-130}$ & 448/394 (1.14) \\
GBM+LAT & LOGP & 0.73$\pm$0.14 & 0.37$\pm$0.04 & -- & -- & 468/398 (1.18) \\
GBM+LAT & Band & 1.34$\pm$0.16 & 2.32$\pm$0.06 & 45$\pm$5 & -- & 485/397 (1.21) \\
GBM+LAT & Band+Cut & 0.6$^{+0.7}_{-1.5}$ & 1.96$^{+0.19}_{-0.08}$ & 26$^{+15}_{-7}$ & 530$^{+400}_{-160}$  & 450/396 (1.14) \\
JOINT & LOGP & 0.81$^{+0.13}_{-0.10}$ & 
               0.35$\pm$0.04 & 
	       -- & 
	       -- &
	       632/544 (1.16) \\	       
JOINT & Band & 1.37$\pm$0.12 & 
               2.31$\pm$0.05 & 
	       45$\pm$4 & 
	       -- &
	       640/543 (1.18) \\
JOINT & Band+Cut & 1.06$^{+0.2}_{-1.18}$ & 
		2.07$^{+0.12}_{-0.08}$ & 
		36$^{+6}_{-12}$ & 
		800$^{+800}_{-300}$ & 
		602/542  (1.11) \\
\hline
\end{tabular}
\label{tabpha1}
\end{table*}


\begin{table*}[!bht]
\centering
\caption{Spectral Fit Results of the time-resolved analysis}
\begin{tabular}{ccccccc}
\hline
\ \ \ \ \ \ Detector \ \ \ \ \ & 
\ \ \ \ \ \ Model \ \ \ \ \ \ &
\ \ \ \ \ \ \ \ \ $-$$\alpha$\ \ \ \ \ \ \ \ &
 \ \ \ \ \ \ \ \ \ $-$$\beta$\ \ \ \ \ \ \ \ & 
 \ \ \ \ \ \ E$_{\rm pk}$ (keV) \ \ \ \ \ \ & 
 \ \ \ \ \ \ \ E$_{\rm cut}$ (keV) \ \ \ \ \ & 
 \ \ \ \ STAT/d.o.f\ \ \ \ \\
\hline
\multicolumn{7}{c}{ Time interval {\it a}: from T$_0$$-$0.3\,s to T$_0$+0.7\,s} \\
\hline
BAT & PL & 1.91$\pm$0.10 & -- & -- & -- & 59/57 (1.03) \\
WAM & PL & 2.42$^{+0.10}_{-0.12}$ &  -- & -- & -- & 39/34 (1.15) \\
WAM & CPL & 1.89$\pm$0.5 &  -- & -- & 600$^{+2000}_{-300}$ &33/33 (1.00) \\
GBM & LOGP & $<$0.017  & 0.54$\pm$0.02 & -- & -- & 313/270 (1.16) \\
GBM & Band & 0.54$\pm$0.16 & 2.27$\pm$0.07 & 56$\pm$5 & -- &  257/269 (0.95) \\
GBM & Band+Cut & 0.4$^{+0.18}_{-0.2}$ & 1.97$\pm$0.17 & 54$\pm$6 & 600$^{+900}_{-200}$ & 247/268 (0.92) \\
GBM+LAT & LOGP & $<$0.019  & 0.55$\pm$0.02 & -- & -- & 314/272 (1.15) \\
GBM+LAT & Band &  0.60$\pm$0.15 & 2.32$\pm$0.06 & 59$\pm$5 & -- &  266/271 (0.98) \\
GBM+LAT & Band+Cut &  0.4$^{+0.18}_{-0.2}$ & 1.97$\pm$0.17 & 54$\pm$6 & 600$^{+900}_{-200}$ & 248/270 (0.92) \\
JOINT & LOGP & $<$0.021 & 
               0.54$\pm$0.02 & 
	       -- & 
	       -- &
	       413/365 (1.13) \\
JOINT & Band & 0.63$\pm$0.13 & 
               2.34$\pm$0.06 & 
	       59$\pm$5 & 
	       -- &
	       369/364 (1.01) \\
JOINT & Band+Cut & 0.41$^{+0.18}_{-0.2}$ & 
		2.02$\pm$0.11 & 
		53$\pm$5 & 
		700$^{+600}_{-300}$ & 
		344/363  (0.95) \\
\hline
\multicolumn{7}{c}{ Time interval {\it b}: from T$_0$+0.8\,s to T$_0$+1.7\,s} \\
\hline
BAT & PL & 1.78$\pm$0.12 & -- & -- & -- & 52/57 (0.92) \\
WAM & PL & 2.38$\pm$0.11 &  -- & -- & -- & 34/34 (1.00) \\
WAM & CPL & 1.8$^{+0.4}_{-0.5}$ &  -- & -- & 1000$^{+4000}_{-600}$ & 28/33 (0.87) \\
GBM & LOGP & 0.35$\pm$0.16  & 0.38$\pm$0.05 & -- & -- & 263/270 (0.97) \\
GBM & Band & 1.22$^{+0.08}_{-0.12}$ & 2.23$^{+0.2}_{-0.13}$ & 140$\pm$30 & -- &  257/269 (0.95) \\
GBM & Band+Cut & 1.22$^{+0.10}_{-0.13}$ & 2.13$^{+0.2}_{-0.13}$ & 140$\pm$30 & $>$900 & 257/268 (0.96) \\
GBM+LAT & LOGP & 0.33$\pm$0.15  & 0.37$\pm$0.05 & -- & -- & 263/272 (0.97) \\
GBM+LAT & Band &  1.31$\pm$0.06 & 2.6$^{+0.8}_{-0.3}$ & 170$\pm$30 & -- &  256/271 (0.94) \\
GBM+LAT & Band+Cut &  1.30$\pm$0.07 & 2.52$\pm$0.17 & 170$\pm$30 & $>$700 & 252/270 (0.93) \\
JOINT & LOGP & 0.29$^{+0.16}_{-0.13}$ & 
               0.38$\pm$0.04 & 
	       -- & 
	       -- &
	       366/364 (1.00) \\	       
JOINT & Band & 1.32$\pm$0.06 & 
               2.51$^{+0.16}_{-0.27}$ & 
	       170$^{+30}_{-20}$ & 
	       -- &
	       361/363 (0.99) \\
JOINT & Band+Cut & 1.29$^{+0.08}_{-0.10}$ & 
		2.34$\pm$0.12 & 
		160$^{+50}_{-20}$ & 
		$>$1000 & 
		356/362  (0.98) \\
\hline
\end{tabular}
\vspace{0.5cm}
\label{tabpha2}
\end{table*}

Additional models, not reported in Table~\ref{tabpha1}, were tested.
A single-temperature black body plus a power-law yields
a poor fit (STAT/d.o.f=905/574), the addition of a high-energy cut-off 
significantly improves the fit (STAT/d.o.f=675/573), but the model
is not statistically preferred to the standard
Band function with a high energy cut-off (STAT/d.o.f=652/573).
A multicolor black body \citep{ryde10} gives similar results.

Table~\ref{tabpha2} reports the results of the
time-resolved spectral analysis for both intervals $a$
and $b$.
As found for the time-integrated spectrum, alternative models 
do not provide an improvement in the fit statistics and are not reported
in the table.

The spectrum of the third peak (interval $c$ in Fig.~1) is well described by a power law
of photon index $\Gamma_{\rm BAT}$=2.78$\pm$0.18. The average observed flux
during this interval is 8$^{+1.1}_{-2.0}$\ee{-7}\,\ergs~in the 15-50\,keV band.

\newpage
\subsection{X-ray data}\label{sec:xray}


The XRT light curve is well described ($\chi^2$/d.o.f.=376/364)
by a power law decay  with slope $\alpha_1$=1.03$\pm$0.04
steepening to $\alpha_2$=1.55$\pm$0.03 at $t_{\rm bk}$=32$^{+9}_{-6}$\,ks.
The two {\it Chandra} detections lie slightly above the
extrapolation of this model, but are consistent with it within 3\,$\sigma$. 
This constrains the time of any late-time jet-break in the X-ray light curve to t$\gtrsim$115\,d.
This time was determined by forcing in the fit an additional break with $\Delta \alpha$=1,
and by varying the break time until a $\Delta \chi^2$=2.706 was reached.

During our observations a slight soft-to-hard spectral evolution 
is visible over the first few hours. We performed time-resolved spectral fits on seven 
consecutive time intervals, selected according to the light curve phases and 
to have $\sim$1000\, net counts each.  
The X-ray spectra were modeled with an absorbed power law.
We derived an intrinsic \nh=9$^{+4}_{-3}$\ee{20}\,\cm{-2}
at $z$=0.49, in excess of the Galactic value of 
2.8\ee{20}\,\cm{-2} \citep{kalberla05}. 
The resulting photon indices $\Gamma_X$, ranging from 2.02$\pm$0.10
to 1.82$\pm$0.09, are consistent within the uncertainties,
however a systematic trend of a slowly decreasing $\Gamma$$_X$ is evident. 
The time-averaged photon index is $\Gamma_X$=1.88$\pm$0.08.


Because of the low number of events in the {\it Chandra} spectrum (67 net counts)
we used the Cash statistics \citep{cash79} and fit it with an absorbed
power law by fixing the absorption components to the values quoted above. 
The resulting photon index is $\Gamma_X$=1.6$\pm$0.3,
from which we calculate an energy conversion factor of $\sim$1.1\ee{-11}\,ergs\,\cm{-2}\,count$^{-1}$.


%


\begin{figure}[!t]
\centering
\vspace{0.3cm}
\includegraphics[scale=0.33,angle=270.]{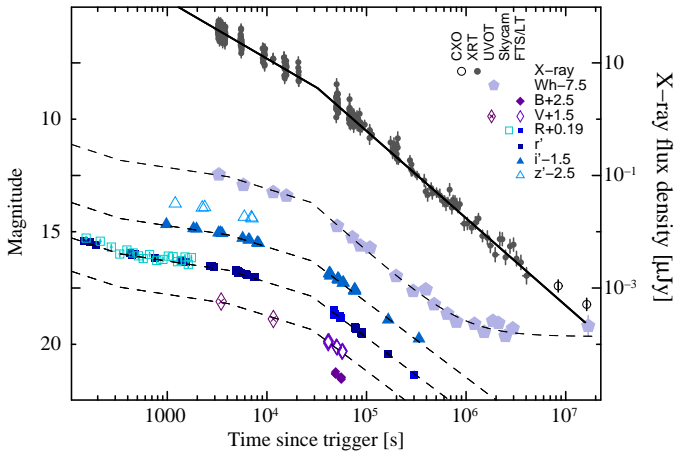}
\caption{X-ray and optical afterglow light curves
of GRB~091127 with the best fit models overplotted
(solid and dashed lines respectively).
At late times (t$>$10\,d) 
the optical emission is dominated by the underlying host galaxy.
Error bars are 1~$\sigma$.
Optical magnitudes are not corrected for Galactic extinction. }
\vspace{0.3cm}
\label{fig:xolc}
\end{figure}


\subsection{Optical data}\label{sec:uvot}

Figure~\ref{fig:xolc}  shows the X-ray  afterglow light curve, reporting the XRT (filled circles) and {\it Chandra} (open circles) data,
and the optical afterglow light curves, including data from UVOT, LT, FTS, and SkycamZ. 
The best fit models are also shown (X-ray: solid line; optical: dashed lines). 

The UVOT/White light curve is well described by a broken power law
plus a constant that accounts for the host galaxy emission.
The afterglow initially decays with a slope of 0.56$\pm$0.04,
steepening to 1.57$\pm$0.05 after $\sim$29~ks. 
We estimate a host galaxy contribution of 23.4$\pm$0.15~mag. 

A significant afterglow color evolution ($\Delta_{I-B}$$\sim$0.25 mag) 
over the course of the first night was reported by Haislip et al. (2009). 
In the fit of the multicolor light curves we initially allowed for frequency-dependent 
slopes and/or temporal breaks, but the sparse sampling 
in the B, V, and z$^{\prime}$ filters does not allow us to detect any color variation. 
As we found consistent results between the different filters, 
we performed a joint fit 
of the BVRr$^{\prime}$i$^{\prime}$z$^{\prime}$ light curves by leaving the normalizations 
free to vary and tying the other model parameters.
The best fit model requires three temporal breaks ($\chi^2$/d.o.f.=53/70).
The model parameters are: $\alpha_1$=0.58$\pm$0.12, $t_{\rm bk,1}$=330$^{+190}_{-70}$\,s, 
$\alpha_2$=0.27$\pm$0.01, $t_{\rm bk,2}$=4.1$^{+0.7}_{-2}$\,ks,
$\alpha_3$=0.55$\pm$0.10, $t_{\rm bk,3}$=28$^{+6}_{-5}$\,ks,
$\alpha_4$=1.34$\pm$0.04.
Contamination from the SN-bump and the host galaxy light, not detected in the 
early-time LT exposures, may explain the shallower temporal index at late times.
By including in the fit a constant component with magnitude
I=22.54$\pm$0.10 to account for the host emission and a SN-like bump,
based on the observation of \citet{cobb10},
the afterglow slope steepens to $\alpha_4$=1.64$\pm$0.06.



\begin{figure}[!t]
\centering
\vspace{0.3cm}
\includegraphics[scale=0.33, angle=270]{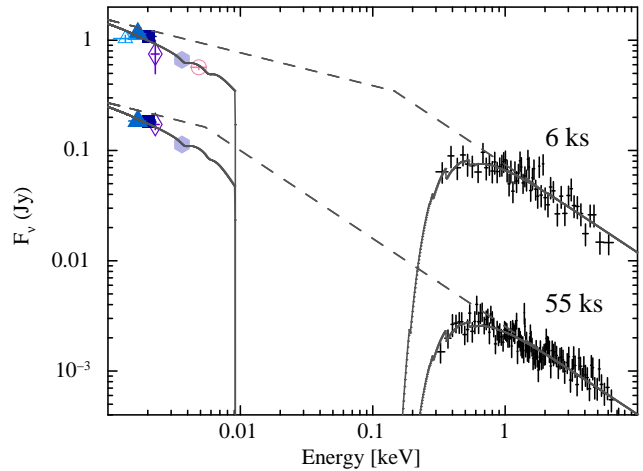}
\caption{Afterglow spectral energy distributions at 6~ks and 55~ks. 
The best fit model (solid line) and the 
same model corrected for extinction and absorption effects (dashed line)
are shown.}
\vspace{0.4cm}
\label{fig:sed}
\end{figure}


\subsection{Spectral energy distribution}\label{sec:sed}

An optical-to-X-ray spectral energy distribution (SED) was produced 
at two different times, 6~ks and 55~ks, selected because of the good color
information and in order to study the spectral evolution across the
achromatic temporal break at $\sim$30~ks. 
Two X-ray spectra were produced, the former in the pre-break interval 9-20~ks, 
the latter in the post-break interval 50-1000 ks, and scaled to match
the observed count-rate at each time of interest. 
The two SEDs were jointly fit in count space \citep{starling07} 
either with a power law 
or a broken power law continuum. In the latter case 
the two spectral slopes were tied so to obey the standard
afterglow closure relations. 
Two dust and gas components, modeling the Galactic and
intrinsic host extinction and absorption, were also included in the fit.
We assumed a Solar metallicity for the absorption components 
and constrained them to the values derived from the XRT spectral fits. 
We tested three canonical laws -- Milky Way (MW), Small Magellanic Cloud (SMC), 
and Large Magellanic Cloud (LMC) -- for the
host galaxy extinction by using the parameterization of \citet{pei92}.

The resulting fit is shown in Figure~\ref{fig:sed}.
Both SEDs are well described ($\chi^2$=146 for 168 d.o.f.) by a broken power law with indices
$\beta_1$=0.300$^{+0.05}_{-0.010}$, $\beta_2$=$\beta_1$+0.5=0.800$^{+0.05}_{-0.010}$ and a decreasing
break energy of $E_{\rm bk}$=0.15$\pm$0.03~keV  at 6~ks and 
$E_{\rm bk}$=6$^{+7}_{-4}$ eV at 55~ks.
A LMC-type extinction with $E(B-V)$=0.036$\pm$0.015~mag is only slightly preferred 
($\Delta \chi^2$$<$2) to a MW-type or a SMC-type law.

\section{Results}\label{sec:res}
\subsection{Prompt emission properties}
\subsubsection{Spectral lags}
A common property of long GRBs is that 
soft energy photons are delayed with respect to the 
higher energy ones. 
The measurement of such lags is a valuable tool
in the study of GRBs and their classification \citep[e.g.][]{gehrels06}.
Systematic studies of BATSE and \swift~ 
bursts show that long GRBs predominantly have
large, positive lags, ranging from 25 ms to $\sim$200 s 
\citep{norris02,norris05,ukwatta10}, while
negligible lags are characteristic of short-duration
bursts \citep{norris06,gehrels06} and high-luminosity long GRBs 
\citep{norris02}.

\begin{figure}[!t]
\centering
\vspace{0.1cm}
\includegraphics[scale=0.35, angle=0]{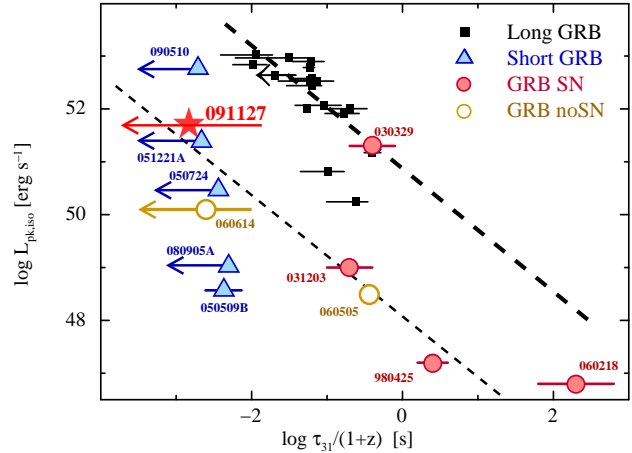}
\caption{Lag-luminosity diagram for long GRBs (squares), 
short GRBs (triangles), nearby GRBs with SNe (filled circles) 
and without SN (open circles).
Error bars are 1~$\sigma$.}
\label{fig:laglum}
\end{figure}

The prompt emission of GRB~091127 seems not to fit in this
classification scheme. 
We measured a small spectral lag of $\tau_{31}$$\sim$2.2~ms, 
consistent with zero, in the BAT channels 3-1, 
and a negative lag of $\tau_{42}$$\sim$$-$9.2~ms in the BAT channels 4-2. 
The burst position in the lag-luminosity plane is shown in Figure~\ref{fig:laglum},
where we also report data for short and
long GRBs from the literature \citep{gehrels06,mcbreen07}.
Having a negligible lag and only a moderate isotropic peak luminosity
($L_{\rm pk,iso}$$\sim$5\ee{51}\lergs), GRB~091127
does not follow the trend of cosmological long GRBs, analogously
to under-luminous bursts such as GRB~980425.  
Nearby sub-energetic bursts (with or without an associated SN)
are outliers of the lag-luminosity relation (thick dashed line).
The inclusion of GRB~091127 suggests that
instead of simply being outliers, there might be a population
of bursts following a distinct trend (thin dashed line). 
While a larger sample of nearby bursts is needed to test this hypothesis, 
an immediate result coming from Figure~\ref{fig:laglum} is that GRB~091127, which is
securely associated with a massive star progenitor, intercepts
the bright end of the short GRB population, showing that the scatter
of long GRBs in the lag-luminosity plane is larger than previously thought.

In the case of GRB~091127, thanks to the GRB low redshift and low intrinsic extinction, 
the associated SN was easily revealed by ground-based follow-up
observations \citep{cobb10,berger11}, nailing down the nature of the GRB progenitor.   
However, had the same GRB occured at a higher redshift, 
its classification would mostly rely on its high-energy properties. 
At $z>3$ the faint soft emission would be under the BAT detection threshold, 
and the GRB would appear as a zero lag, intrinsically short 
($T_{90}/(1+z) \lesssim$2~s) burst, similar to GRB~080913 and GRB~090423
for which a merger-type progenitor was also considered \citep[e.g.][]{zhang09}.
It is also possible that some of the higher redshift short-duration bursts 
arise from massive star collapses \citep[e.g.][]{virgili11}.




\subsubsection{Softening of the high-energy spectrum}

\begin{figure}[!t]
\centering
\vspace{0.2cm}
\includegraphics[scale=0.35,angle=270.]{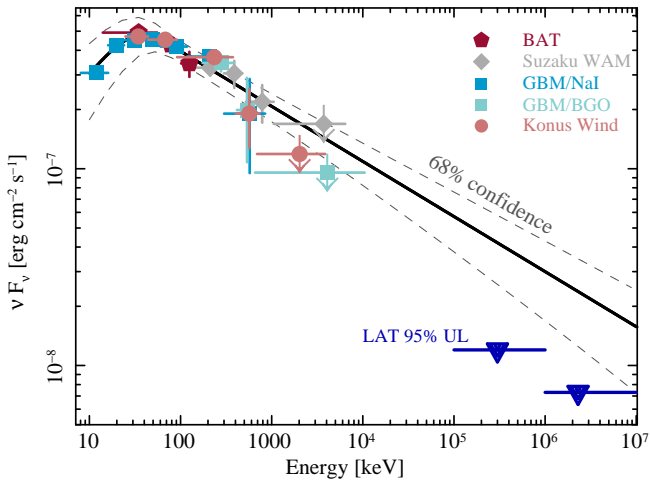}
\caption{Best-fit Band model of the time-averaged spectrum (solid line) 
with its 1~$\sigma$ confidence interval (dashed lines). 
Data from \swift/BAT, {\it Suzaku}-WAM, GBM and Konus-{\it Wind} are reported
with their 1~$\sigma$ error bars.
Upper limits from \fermi/LAT are also shown.}
\vspace{0.2cm}
\label{sedlat}
\end{figure}

Fitting results are listed in Table~1 for the time-averaged spectrum
and in Tab.~2 for the time-resolved analysis.
By describing the time-integrated spectrum with the canonical Band function 
we obtained typical parameters: $\alpha$$\sim$$-$1.3, $\beta$$\sim$$-$2.3 and a soft peak energy of $\sim$45\,keV.
However by extrapolating the best fit Band model to the LAT energy range,
the predicted flux in the 100\,MeV\,--\,1\,GeV energy band is $\approx$\e{-7}\ergs,
well above the 95\% upper limit derived in \S~\ref{sec:lat}.
This is shown in Figure~\ref{sedlat}, where we report the 
observed data with their best fit Band model
extrapolated to the LAT energy range.

The joint fits reported in Table~1 confirm that \fermi/LAT~observations 
are not consistent with the extension of a Band function 
from low to high energies, but require a steepening of the 
spectrum at energies below 100 MeV.
The inclusion of a high-energy spectral break,
that we modelled as an exponential cut-off,
improves the fit ($\Delta$-STAT$=$38 for one additional degree of freedom).
Such a break is particularly evident in the {\it Konus-Wind} 
and in the GBM spectra, and we note that the two fits
yield consistent values of the cut-off energy and an
improvement in the fit statistic of $\Delta \chi^2$=14
and $\Delta \chi^2$=9 respectively.
The quality of the data does not allow us to constrain
the spectral index above the break energy and 
distinguish between a steepening of the power-law decay
or an exponential cut-off. By modeling the high-energy data
with a simple power-law we derive a photon index of $\sim$$-$3.6,
and set an upper limit $<$$-$2.6 (90\% confidence level). 
The significance of the high-energy break was tested 
by simulating 10,000 spectra with a simple Band shape. 
We jointly fit each set of spectra with a Band function 
(our null model) and a Band function with an exponential cut-off 
(the alternative model). 
The fractional number of simulations in which $\Delta$-STAT$\geq$38
gives the chance probability that a high-energy spectral break 
improves the fit.
None of the simulations showed a variation of the statistics
as high as the one observed, 
confirming that the presence of a spectral break is 
statistically preferred at a $>$99.99\% level.

The log-parabolic model of \citet{massaro10} also provides a better fit than the standard
Band function (STAT/d.o.f=679/575 vs. 690/574), and naturally accounts
for the observed suppression of the high-energy emission.

A time-resolved spectral analysis temporally  
localizes the spectral break during the first 
$\gamma$-ray peak (interval $a$).
In this case the presence of a cut-off at energies $\approx$500-1000\,keV
decreases the fit statistics of $\Delta$-STAT=25. 
The lower significance with respect to the
time-averaged analysis is likely due to the lack of {\it Konus-Wind} data in this fit,
however the observed break is evident both in the WAM and in the GBM spectra
at a folding energy E$_{\rm cut}$ consistent between the different instruments.
According to this model, the observed fluence during the first peak
is (4.3$\pm$0.6)\ee{-6}\,erg\,\cm{-2} in the 8-1000\,keV energy band.
At a redshift $z$=0.49 this corresponds to an isotropic equivalent energy 
$E_{\gamma,\rm iso}$=(3.5$\pm$0.5)\ee{51}\,erg in the 1-10,000 keV rest-frame energy band.
In this time interval the derived value of the low-energy 
index is $\alpha$=$-$0.41$^{+0.18}_{-0.2}$, which is harder but
marginally consistent with the 
limit of 2/3 imposed by the optically thin synchrotron emission.
The presence of a thermal component is sometimes
invoked to explain the hardest low-energy spectral indices
\citep[e.g.][]{ghirlanda03}.
As already noted in \S~\ref{sec:gamma}, 
we tested this hypothesis and found that in no case does the inclusion
of a black-body (single or multi-temperature) yield a significant
improvement in the fit statistics, although 
such a component is not inconsistent with the data.

The spectrum of the second peak (interval $b$) can be well described by a Band function. 
The inclusion of the LAT data yields a steeper high-energy spectral slope
than the one derived from the GBM only fit,
and the addition of a high-energy  break is not required by the data. 
According to this model, the observed fluence during this interval
is (4.5$\pm$0.2)\ee{-6}\,erg\,\cm{-2} in the 8-1000\,keV energy band, corresponding to an isotropic equivalent energy 
$E_{\gamma, \rm iso}$=(4.3$\pm$0.3)\ee{51}\,erg in the 1-10,000 keV rest-frame energy band.


\begin{figure*}[!t]
\centering
\includegraphics[scale=0.53, angle=0]{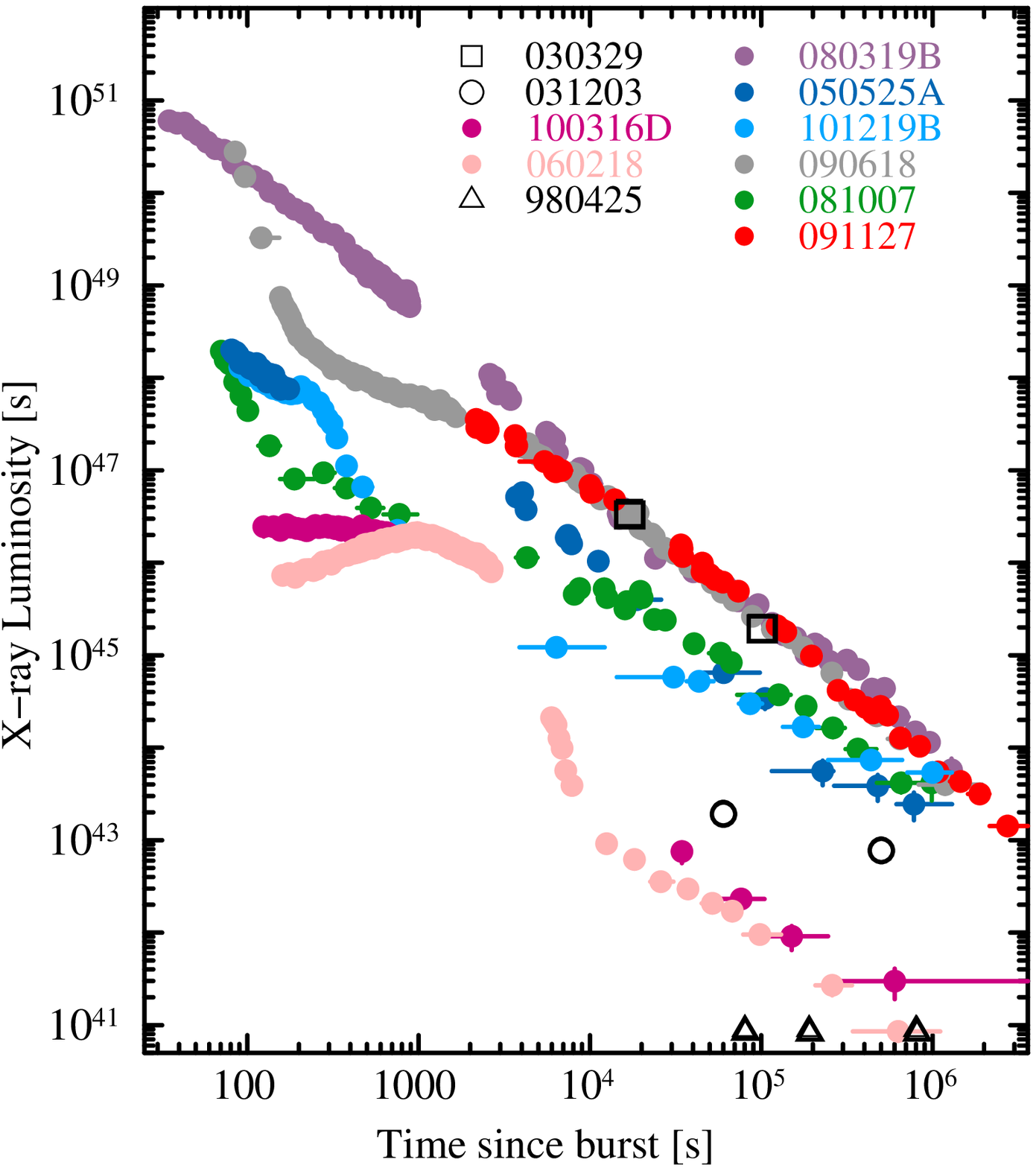}
\hspace{0.6cm}
\includegraphics[scale=0.543, angle=0]{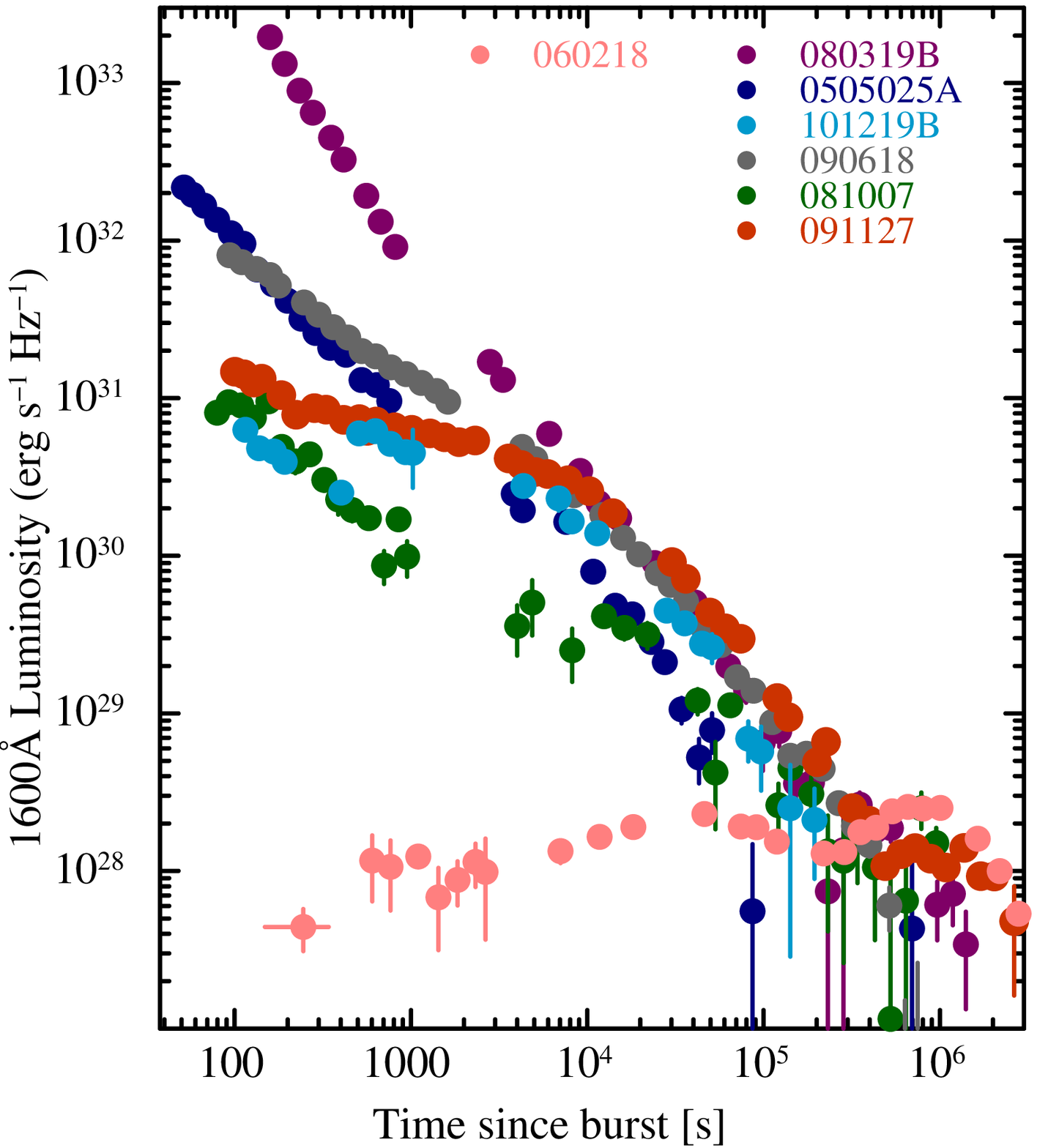}
\caption{{\it Left panel:} Rest-frame XRT afterglow lightcurves
for {\it Swift} GRBs with an associated SN (filled circles). 
We also report the data
for the three pre-\swift~bursts with a spectroscopically confirmed SN
(open symbols, \citealt{kaneko07}). 
{\it Right Panel:} Rest-frame UVOT afterglow light curves. 
Only GRBs with an UVOT detection are shown. 
Early time LT/FTS data for GRB~091127 are also reported. }
\vspace{0.3cm}
\label{fig:ag}
\end{figure*}



\subsection{Afterglow properties}\label{sec:ag}
 


In Figure~\ref{fig:ag} we compare the afterglow of GRB~091127 to the sample of 
\swift~GRBs with bona fide SN associations \citep{hbloom11}.
The observed XRT and UVOT 
light curves were corrected for redshift and absorption effects,
and shifted to a common rest-frame energy band of 0.3-10~keV (XRT) and 
a rest-frame wavelength of 1600\,\AA\ (UVOT; \citealt{oates09}).
From an afterglow perspective, 
GRB~091127 resembles the behavior of typical long GRBs,
dominated by the bright emission from the external forward shock,
rather than the unusual evolution of nearby GRBs. 
The isotropic X-ray luminosity at t=11~hr is $L_{X,\rm iso}$$\sim$2\ee{45}\,erg\,s$^{-1}$,
very similar to GRB~030329, and a factor of $>$\e{3} brighter than 
GRB~031203 and other GRBs/SNe. 
The UV/optical afterglows appear instead to decay more rapidly and 
to cluster at late times, but 
this could be the result of an observational bias, as
the chance of discovering a supernova is higher if the 
optical afterglow is faint.

If the afterglow emission of GRB~091127 is mainly synchrotron 
radiation from the external forward shock, its broadband behavior 
has to obey the fireball model closure relations \citep[e.g.][]{zhames04}.
We found that the GRB afterglow is roughly consistent with a
model of a narrow jet expanding into a homogeneous surrounding medium.
Our results agree well with previous studies \citep{vergani11,filgas11}.
The fireball model describes the emission from a population of accelerated electrons with energy
distribution $n(\epsilon)$\,$\propto$\,$\epsilon^{-p}$.
From the afterglow spectral properties we derive an electron 
index $p$=1.60$^{+0.10}_{-0.02}$, which is at the lower end of the $p$ distribution but not uncommon \citep{pk02}. 
An achromatic break is detected at $\sim$8\,hr,
after which the X-ray and optical light curves decay with a similar
slope of $\sim$1.6.
This behavior is suggestive of an early jet-break.
The presence of a jet-break at early times is also supported by 
our {\it Chandra} observations, which do not show
evidence of a steepening in the X-ray light curve
several months after the burst.
We found that any possible late time jet-break is constrained to t$>$115\,d,
which, for typical parameters, would imply an unusually large opening
angle $\theta_j$$>$30$^{\circ}$.
Instead the two {\it Chandra} points hint at a shallower decline, as expected
for example in the transition to the non-relativistic phase \citep{piro01}.

The SED analysis (\S~\ref{sec:sed}) shows that optical and X-ray data belong to
different branches of the synchrotron spectrum, since
the cooling frequency $\nu_c$ lies between the two energy bands.
The observed break at 30~ks is therefore not connected to spectral variations
or changes in the ambient density. 
Figure~\ref{fig:sed} shows that at 6~ks the lowest optical  flux produced
by the X-ray source (with $\nu_c$ just below X-rays) would be only a factor $\lesssim$2 
lower than measured, thus the reverse shock contribution to the total optical flux 
\citep{koba00} is negligible and optical and X-ray 
emission mainly arise from the
same source (external forward shock).
In this framework, the evolution of the cooling frequency
is tied to the observed X-ray and optical decays by 
the following relation \citep{p06}: \\
\begin{equation}
- \frac{d~{\rm ln\,} \nu_c}{d~{\rm ln\,} t} = 2 (\alpha_X -\alpha_{\rm opt}) = 0.94 \pm 0.11.
\end{equation}\\
This is consistent with our spectral fits, which measure a cooling frequency  
that is rapidly moving downwards in energy as $\nu_{c} \propto t^{-1.5\pm0.5}$, 
as independently found in \citet{filgas11}. 
For constant microphysical parameters, a decreasing $\nu_c$ suggests 
an ISM environment
rather than a wind-like density profile, where the cooling frequency is
expected to increase with time. 
In a uniform density medium, the cooling break evolves
as $\nu_c\propto E^{-1/2} \epsilon_B^{-3/2} t^{-1/2}$ 
for a spherical expansion,
and as $\nu_c \propto E^{-2/3} \epsilon_B^{-3/2} t^0$ in the jet spreading phase
\citep{pk02,pk04}.
For constant microphysical parameters and no energy injection into the blastwave,
the expected decay is shallower than the observed one.
This shows that the simplest version of the fireball model 
can not account for the overall afterglow behavior, 
and, as we will discuss in Sect.~\ref{energy}, some modifications (e.g. energy injection or evolving microphyscal parameters) are required.





\section{Discussion}\label{sec:discuss}
\subsection{Origin of the high-energy spectral break}

The most recent \fermi~observations of GRBs suggested that the prompt
$\gamma$-ray emission can be satisfactorily described
by a smoothly broken power law, the Band function, extending to 
the GeV energies, 
often accompanied by an additional non-thermal component
modeled as a power law \citep{binbin11}.
In this burst we found that a standard Band function, 
though providing an adequate description of the
spectrum in the keV energy range, is in contrast with the
simultaneous \fermi/LAT observations as it overpredicts
the observed emission above 100\,MeV (see Fig.~\ref{sedlat}).
The spectral fits presented in Table~1 and Table 2 detect
the presence of a spectral softening at $\approx$0.5-1\,MeV
in the time-integrated spectrum and during the first peak
of emission. 
This disfavors spectral evolution as the origin of the
observed feature.

A steepening of the high-energy spectral slope could be caused by several factors
such as absorption from the extragalactic background light (EBL), 
attenuation via pair production ($\gamma\gamma \rightarrow e^{\pm}$) or 
an intrinsic break in the energy distribution 
of the emitting electrons. 
Based on the low redshift of this burst ($z$=0.49) and the low energy of the
observed break, EBL absorption can be excluded \citep[see e.g.][]{finke10}. 
Below we consider in turn the other possibilities.

\textit{Optical depth effects --}
The lack of high-energy photons in bright bursts such as GRB~091127
could be an indication of a pair opacity break \citep{beniamini11}, 
and therefore used  to constrain the outflow Lorentz factor \citep{litsari01}. 
In order to be self-consistent these calculations rely on the fundamental 
assumption that the observed sub-MeV spectrum extrapolates to GeV energies.
Following this line of argument, we can set a first upper limit on the 
bulk Lorentz factor in GRB~091127 just by considering its non-detection by LAT.
We use here the Band function parameters and impose $E_{\rm max}$$<$100\,MeV:\\
\begin{equation}
\Gamma_{\gamma\gamma} <
130 \left[\left(\frac{E_{\rm max}}{100 {\rm \ MeV}}\right)\,f_{100}\,\,t_v^{-1}\right]^{\frac{1}{2\beta+2}},
\end{equation}\\
where $\beta$=2.28 is the high-energy spectral slope and
$f_{100}\sim$0.1 ph\,\cm{-2}\,s$^{-1}$\,keV$^{-1}$ the observed flux density at 100 keV, 
both derived from the spectral fit in Tab.~1. The
variability timescale was set to t$_v$$\approx$0.3\,s, 
the minimum value observed in the $\gamma$-ray lightcurve.
In deriving Eq.~2 we approximated the spectral shape with a simple
power-law, $f_{\nu}\propto \nu^{-\beta}$. Given that for $E$=100\,MeV 
the typical energy of the target photons is \mbox{ $E_{t}\sim$1 MeV $>>$ $E_{\rm pk}$},
the effects of low-energy spectral curvature \citep{baring98}  can be considered
negligible and a simple power-law decay is a valid approximation.

The upper limit derived in Eq.~2 is based on the simple formulation given in
\citet{litsari01}, where spatial and temporal dependencies are averaged out. 
More realistic calculations taking into
account the progressive buildup of the radiation field 
further decrease the above value by a factor of 2-3 
\citep{hascoet11}, that is 
$\Gamma_{\gamma\gamma}$\,$\approx$\,50.
This is significantly
lower than the values estimated for cosmological GRBs 
\citep{molinari07,liang11}, though similar 
to the Lorentz factor inferred for X-ray flares \citep{100728a}.  
If we now take into account the observed steepening at $\lesssim$1 MeV
as it originates from an increase in the optical depth, 
by setting $E_{\rm max}\simeq$$E_{\rm cut}$
we get  $\Gamma$\,$\approx$\,2. 
Such a low Lorentz factor, though atypical for classical GRBs, 
is not unprecedented \citep{sode06}. 
A weakly relativistic outflow could therefore account for the
lack of high-energy photons, and the observed soft spectrum, 
but not for the bright afterglow detected a few minutes after the burst.

An independent estimate of the bulk Lorentz factor can be derived from
afterglow observations. 
The duration of the GRB being rather short, we consider the thin
shell case \citep{koba99}.
Since the afterglow is already fading in our first observation we can assume
that the onset happened at $t_{\rm pk}<$140\,(1+$z$)$^{-1}$ $\sim$100\,s, and set a lower limit
to the outflow Lorentz factor $\Gamma_0$ \citep{piran99}:
\begin{equation}
\Gamma_0 > 240~\left( \frac{E_{\gamma,52}}{\eta_{0.2}~n~t^3_{\rm pk,2}} \right)^{1/8} ,
\end{equation}
where E$_{\gamma}$=\e{52} E$_{\gamma,52}$~erg is the isotropic-equivalent
energy, $\eta$ = 0.2$\eta_{0.2}$ is the radiative efficiency and $n\sim$1\,\cm{-3} is the medium density \citep{bloom03}. 
By using the empirical relation suggested by \citet{liang11}, we 
infer a similar high value of $\Gamma_0 \sim$200.

The limits derived from the prompt and afterglow emission
properties are inconsistent: the former suggest a mildly
relativistic outflow ($\Gamma<50$, or even $\Gamma\approx$2), the latter a highly relativistic jet ($\Gamma>>100$).
A possibility that would reconcile the two sets of limits is that the first spectrally softer pulse, 
during which we detect the significant presence of a spectral break, is instead the GRB precursor 
originating at $R$\,$\approx$$2\Gamma^2 c t_v$$\approx$\,\e{11} cm, e.g. from the jet cocoon emerging from the progenitor star \citep{lazzati05}. 
A different physical origin could also explain the different lags between
the two main $\gamma$-ray events and the unusual lag evolution:
while spectral lags in GRB pulses generally tend to increase with time \citep{hakkila08}, 
it has been found that precursors have larger lags than 
the following $\gamma$-ray emission \citep{page07}.
Precursors, however, carry only a small fraction of the total energy release \citep{morsony07}, 
while the first peak encloses 50\% of the observed $\gamma$-ray fluence.

We therefore  are led to consider that our assumption of 
a pair opacity break is not valid, that is: 
1) the inconsistency between Eq.~2 and Eq.~3
implies that the Band-type spectrum does not extend 
to GeV energies, but a spectral break 
(not related to optical depth effects) 
below 100 MeV is required by the data;
2) we identify this break with the steepening at $\approx$0.7 MeV, 
which is therefore an intrinsic feature 
of the GRB spectrum. \\



\textit{Breaks in GRB spectra --}
We discuss here the standard scenario, in which
internal shocks within the expanding outflow 
accelerate the ambient electrons to relativistic energies
with a power-law distribution $ n (\epsilon) \propto \epsilon^{-p}$.
The GRB prompt emission originates as synchrotron radiation 
from the shock-accelerated electrons.
The small ratio between the GeV and keV fluences of this burst, 
$\mathcal{F}_{\rm GeV}/\mathcal{F}_{\rm keV}\lesssim0.01$, disfavors 
Synchrotron Self Compton as the main radiation mechanism 
\citep{piran09,beniamini11}. 

The observed spectrum  of the first $\gamma$-ray peak, 
 is roughly in agreement with a 
fast cooling synchrotron spectrum:
the low-energy slope $\alpha$$\sim$$-$0.4 is marginally consistent with the
maximum slope of 2/3 allowed for $\nu$$<$$\nu_c$, while
the high-energy slope $\beta\sim$$-$2 suggests that for $E$\,$>$50~keV we are 
already above the injection frequency $\nu_m$. 
It follows that $\nu_c$$\approx$$\nu_m$, that is the effects of 
adiabatic and radiative cooling are comparable 
(marginal fast cooling; \citealt{daigne11}).
In the extreme case $\Gamma_c/\Gamma_m \sim$10, synthetic spectra 
resemble the observed spectral shape: 
a hard low-energy tail followed by a smooth, flat transition ($\nu_m$$<$$\nu$$<$10~$\nu_m$)
to the final $F_{\nu} \propto \nu^{-p/2}$ decay.
The observed steepening at $\approx$0.7~MeV from $\beta$$\sim$$-$2 to $<$$-$2.6 corresponds 
to this transition, and implies $p\gtrsim3.2$.
However when the slow cooling contribution is significant,
the radiative efficiency decreases markedly \citep{daigne11}, 
and it is hard
to account for the high luminosity and variability of the prompt emission.
If we consider the more efficient case of
$\Gamma_c/\Gamma_m \sim$1, then the
spectral break has to be ascribed to a different mechanism.

A spectral cut-off is expected at $\nu(\gamma_M)$, where $\gamma_M$
is the maximum Lorentz factor of the shocked electrons. 
Such a break occurs at energies $\gtrsim$200~MeV \citep{bosnjak09}, 
and it is unlikely at the origin of the MeV break. 
An alternative explanation is an intrinsic curvature in the 
energy distribution of the radiating electrons \citep{massaro10}, 
arising if the higher energy electrons are 
accelerated less efficiently than those with lower energy. 

\subsection{Jet collimation and energetics}\label{energy}

From our broadband spectral fits of the prompt emission
we derived an isotropic equivalent energy $E_{\gamma,\rm iso}=$(1.1$\pm$0.2)\ee{52}\,erg,
which is in the typical range of long GRBs \citep{bloom03}.
The afterglow properties show evidence of a tightly collimated outflow,
indicating that the true energy release is significantly lower. 
The achromatic nature of the break at $t_{\rm bk} \sim$30~ks and the subsequent 
afterglow fast decay are typical signatures of a jet-break,
and we first consider this hypothesis.
In this scenario the jet opening angle $\theta_j$ is:
\begin{equation}
\theta_j = 4.2 \left(\frac{E_{\rm iso,52}}{\eta_{0.2} n}\right)^{-1/8} 
\left(\frac{t_{\rm bk}}{8 \rm~hr}\right)^{3/8}\,{\rm deg},
\end{equation}
and the collimation-corrected energy is E$_{\gamma,j}$=(3.0$\pm$0.8)\ee{49}\,erg, 
at least an order of magnitude lower than typical long GRBs \citep{cenko10}.
However, as noted in \S~\ref{sec:ag}, this simple fireball 
scenario fails to reproduce two main features: 1) the rapid temporal evolution
of the cooling frequency; 2) the 
observed pre-break flux decay rates ($\alpha_X$ = 1.03$\pm$0.04, $\alpha_{\rm opt}$ = 0.56$\pm$0.04), which are not compatible with the model expectations
($\alpha_{\nu>\nu_c}\sim$0.7, $\alpha_{\rm \nu_c>\nu>\nu_m}\sim$0.45
for a spreading jet; $\alpha_{\nu>\nu_c}=\alpha_{\rm \nu_c>\nu>\nu_m} \sim$0.8 for a non-spreading jet; \citealt{pk04}).
In order to reconcile the observed afterglow behavior
with the theoretical expectations one needs to invoke either a
continual energy injection and/or evolving microphysical
parameters. 
The former scenario would require an extreme injection episode, 
the jet energy increasing by a factor of 100 in the first 8 hours.
Furthermore, there is no apparent reason for the injection
to end at the time of the jet-break, 
leading to an even larger shock energy carried by the slower ejecta.
The alternative possibility of a growing magnetic energy fraction $\epsilon_B$
is discussed by \citet{filgas11}.


\begin{figure}[!t]
\centering
\vspace{0.2cm}
\includegraphics[scale=0.33]{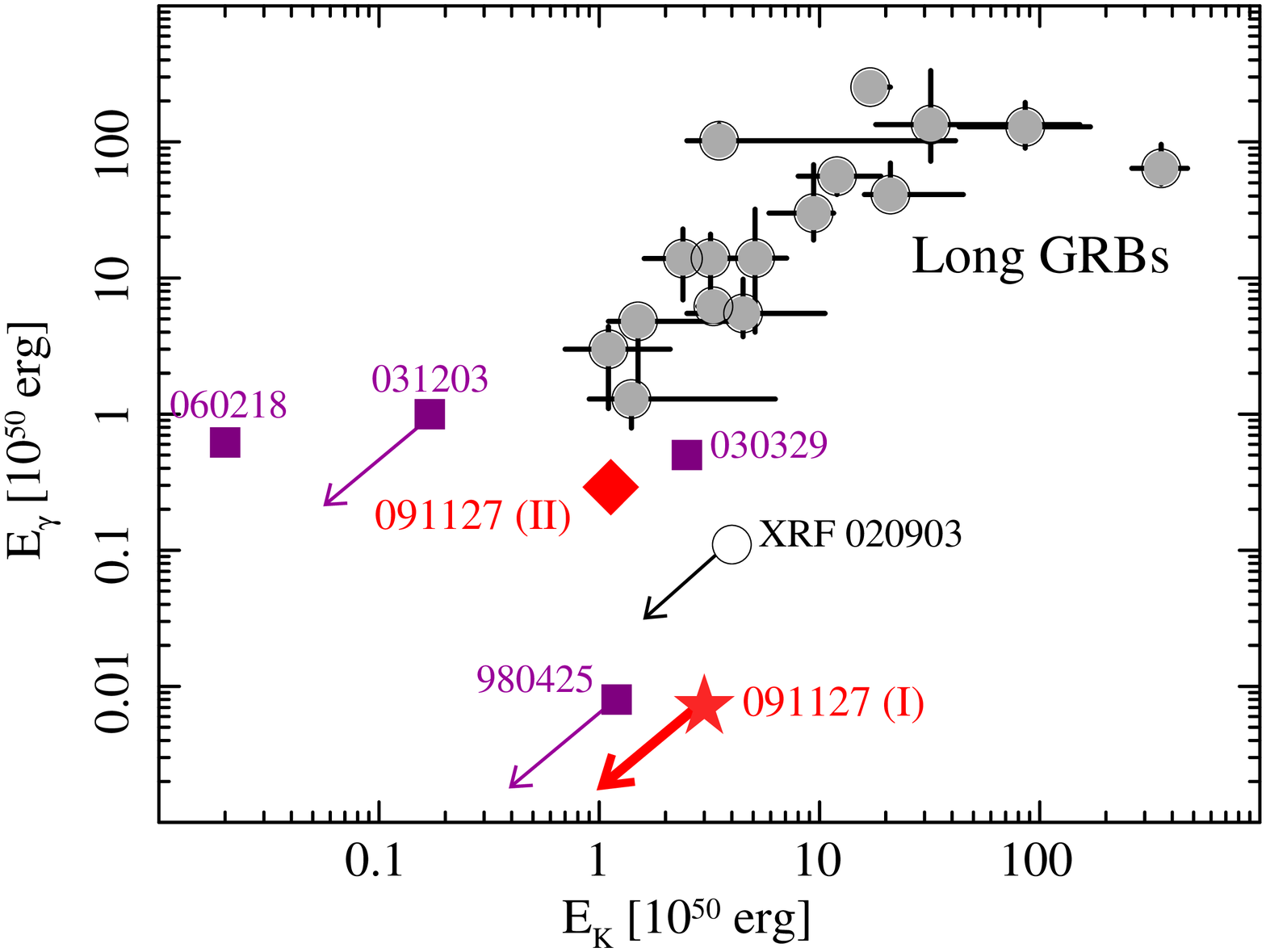}
\caption{Prompt emission energy release, $E_{\gamma}$, versus
afterglow kinetic energy, $E_{K}$. 
For GRB~091127 both the scenarios discussed in the text
are shown: (I) narrow jet + prolonged
energy injection; (II) evolving $\epsilon_B$.
We report data for standard long GRBs (filled circles), 
XRFs (open circles), and bursts with a spectroscopic SN (squares). 
Values are corrected for collimation effects.
Ref: \citet{pk02,bloom03,soderberg04,cenko10}.}
\vspace{0.2cm}
\label{fig:erel}
\end{figure}


We found instead that a narrow confined jet,
whose boundary is visible from the first afterglow measurement 
(i.e. $\Gamma$ $<$ $\theta_j^{-1}$), and a prolonged energy injection, 
lasting until $\sim$30 ks,
provide a consistent description of the afterglow temporal 
and spectral properties and ease the energetic burden
without requiring any variation of the shock microphysical parameters. 
For an ISM-like circumburst medium (\S~\ref{sec:ag}), 
the flux decay indices are given by \citep{pk04}:
\begin{eqnarray}
\alpha_{o} &=& \frac{3}{4}p -\frac{p+4}{4}e,  \\
\alpha_{X} &=&  \frac{3p+1}{4} - \frac{p+3}{4} e,  
\end{eqnarray}
where $e$ is the power-law evolution of the forward-shock energy $E 
\propto t^{e}$. The above set of equations overconstrain the $e$ parameter, thus
providing a consistency check of the solution.
By substituting in Eq.~5 and 6 the observed pre-break temporal slopes 
and the value of $p$$\sim$1.6 from the broadband spectral fit, we derive 
$e$=0.48$\pm$0.06 and $e$=0.39$\pm$0.06, respectively.
Departures of the energy injection from a pure power law can 
explain the optical plateau at t$<$5~ks, while the
cessation of energy injection at $\approx$30 ks yields the
observed achromatic break.
According to this model, by imposing $t_{\rm bk}<140$\,s in our Eq.~4 we derive 
$\theta_j \lesssim$0.6~($n$/1\cm{-3})$^{1/8}$ deg, 
and $E_{\gamma} \lesssim$6\ee{47} ($n$/1\cm{-3})$^{1/4}$\,erg.
By using $e\sim$0.45, 
the blastwave kinetic energy can be constrained to $E_{\rm K}\lesssim$3\ee{50}\,erg,  
most of which comes from the slower ejecta that are 
gradually replenishing the forward shock energy. 

From our analysis the following features clearly emerge:
GRB~091127 is characterized by a highly collimated outflow ($\theta_j$$\lesssim$4$^{\circ}$),
a low prompt $\gamma$-ray energy ($E_{\gamma}$\,$<$3\ee{49}\,erg),
and a total relativistic energy yield of $E_{\rm rel}$\,$\lesssim$3\ee{50}\,erg,
at the lower end  of the long GRBs distribution.
In Figure~7 we compare the burst energetics with the sample of long
GRBs. Independently from the afterglow model adopted 
(I, narrow jet + energy injection: star;
II, evolving $\epsilon_B$: diamond), the burst location in the lower left corner shows that GRB~091127 more closely resembles the class of X-ray Flashes (XRFs) and GRBs/SNe rather than 
typical GRBs. This is also consistent with its rather soft spectrum and unusual
lags.

\section{Summary and Conclusion}\label{sec:end}

We presented a broadband analysis of the prompt and afterglow emission of GRB~091127, 
securely associated with SN2009nz. 
Two main features emerged from our study of the prompt emission:
1) the burst is characterized by small, negligible spectral lags;
2) the high energy ($>$100\,MeV) emission is significantly suppressed. 
The GRB has a long duration ($T_{90}$$\sim$7\,s), 
and a relatively soft spectrum ($E_{\rm pk}$$\approx$45\,keV).
However having negligible spectral lags 
and only a moderate luminosity, 
the burst does not fit the lag-luminosity relation followed by 
cosmological long GRBs, but lies in the region of short
duration bursts. While the association with SN2009nz leaves no doubts
about the origin of the GRB progenitor, the atypical lag behavior adds additional uncertainty
in the classification of GRBs based solely on their high-energy properties.
It also links GRB~091127 to nearby sub-energetic bursts,
such as GRB~980425, which are also outliers of the lag-luminosity relation. 

By modeling the GRB prompt emission with the standard Band function, 
we found that such a model significantly overpredicts the 
observed flux at higher ($>$100~MeV) energies. 
Consistently, our spectral fits show evidence 
of a spectral curvature at energies $\lesssim$1~MeV. 
If due to opacity effects, the suppression of high energy emission would 
suggest a low outflow Lorentz factor ($\Gamma$\,$<$\,50, or even $\Gamma$\,$\approx$\,2), 
as measured in nearby sub-energetic GRBs. 
However, this interpretation is not consistent with our early-time detection of a bright
fading afterglow, which suggests $\Gamma$$>>$100. We therefore 
conclude that the high-energy break is an intrinsic property of the GRB spectrum.

The multi-wavelength afterglow emission is characterized by an achromatic break 
at $\sim$8~hr after the burst, and by a rapidly decaying cooling frequency, $\nu_c$\,$\propto$\,t$^{-1.5 \pm 0.5}$. 
We considered two scenarios to interpret these features within the standard fireball model. 
The former interprets the achromatic break as a jet-break, from which  we
derive a jet opening angle $\theta_j$$\approx$4\,deg, and a collimation-corrected energy $E_{\gamma}$$\approx$3$\times$\e{49}\,erg. This model needs to let 
the microphysical parameters vary with time in order 
to reproduce the observed temporal decays and the 
rapidly decreasing $\nu_c$.
The latter scenario instead interprets the achromatic break as the end of a prolonged
energy injection episode, the jet-break happening before the start 
of our observations (t$<$140\,s). 
According to this model, we derive a jet opening angle $\theta_j$\,$\lesssim$\,0.6\,deg, and a collimation-corrected energy $E_{\gamma}$\,$\lesssim$\,6$\times$\e{47}\,erg.
This GRB therefore presents hybrid properties: a high luminosity $\gamma$-ray emission
powered by narrowly collimated and highly relativistic outflow as typical of long GRBs; 
its low-energy output, rather soft spectrum and location in the lag-luminosity plan
more closely resembles the class of XRFs and GRBs/SNe.


\acknowledgements{}

We thank T. Ukwatta, R. Starling, and 
and R. Filgas for useful discussions and information.

ET was supported by an appointment to the NASA Postdoctoral Program at the Goddard Space Flight Center, administered by Oak Ridge Associated Universities through a contract with NASA. 

This work made use of data supplied by the UK Swift Science Data Centre at the University of Leicester.

The Konus-Wind experiment is supported by a Russian Space Agency contract and RFBR grant 11-02-12082-ofi\_m.

The \textit{Fermi} LAT Collaboration acknowledges generous ongoing support
from a number of agencies and institutes that have supported both the
development and the operation of the LAT as well as scientific data analysis.
These include the National Aeronautics and Space Administration and the
Department of Energy in the United States, the Commissariat \`a l'Energie Atomique
and the Centre National de la Recherche Scientifique / Institut National de Physique
Nucl\'eaire et de Physique des Particules in France, the Agenzia Spaziale Italiana
and the Istituto Nazionale di Fisica Nucleare in Italy, the Ministry of Education,
Culture, Sports, Science and Technology (MEXT), High Energy Accelerator Research
Organization (KEK) and Japan Aerospace Exploration Agency (JAXA) in Japan, and
the K.~A.~Wallenberg Foundation, the Swedish Research Council and the
Swedish National Space Board in Sweden.

Additional support for science analysis during the operations phase is gratefully
acknowledged from the Istituto Nazionale di Astrofisica in Italy and the Centre National d'\'Etudes Spatiales in France.


\bibliographystyle{aa}
\bibliography{ref091127}

\end{document}